\documentclass[aps,preprint,superscriptaddress,preprintnumber]{revtex4}
\def\be{\begin{equation}}
\def\ee{\end{equation}}
\def\bea{\begin{eqnarray}}
\def\eea{\end{eqnarray}}
\usepackage{graphicx}
\def\met{\slash{\!\!\!\!E}_T}
\begin{document}

\preprint{ANL-HEP-PR-11-14, EFI-11-6, UCRHEP-T500}
\title{Probing Lepton Flavor Triality with Higgs Boson Decay}

\author{Qing-Hong Cao}
\affiliation{High Energy Physics Division, Argonne National Laboratory, 
Argonne, IL 60439, USA}
\affiliation{Enrico Fermi Institute, University of Chicago, Chicago, 
IL 60637, USA}

\author{Asan Damanik}
\affiliation{Department of Physics, Sanata Dharma University, Yogyakarta, 
Indonesia}

\author{Ernest Ma}
\affiliation{Department of Physics and Astronomy, University of California, 
Riverside, California 92521, USA}

\author{Daniel Wegman}
\affiliation{Department of Physics and Astronomy, University of California, 
Riverside, California 92521, USA}

\begin{abstract}
If neutrino tribimaximal mixing is explained by a non-Abelian discrete 
symmetry such as $A_4$, $T_7$, $\Delta(27)$, etc., the charged-lepton 
Higgs sector has a $Z_3$ residual symmetry (lepton flavor triality), 
which may be observed directly in the decay chain $H^0 \to \psi_2^0 
\bar{\psi}_2^0$, then $\psi_2^0(\bar{\psi}_2^0) \to l_i^+ l_j^- ~(i \neq j)$, 
where $H^0$ is a standard-model-like Higgs boson and $\psi_2^0$ is a 
scalar particle needed for realizing the original discrete symmetry.  
If kinematically allowed, this unusual and easily detectable decay is 
observable at the LHC with 1 fb$^{-1}$ for $E_{\rm cm} = 7$ TeV.
\end{abstract}

\maketitle

\section{Introduction}

In recent years, a theoretical understanding of the observed pattern of 
neutrino mixing, i.e. the $3 \times 3$ matrix $U_{l \nu}$ which links 
charged-lepton mass eigenstates to neutrino mass eigenstates, has been 
achieved in terms of non-Abelian discrete symmetries.  
In particular, 
the tetrahedral symmetry $A_4$~\cite{mr01} has been shown to be 
successful~\cite{m04} in explaining tribimaximal mixing~\cite{hps02}, i.e.
\begin{equation}
U_{l \nu} = \pmatrix{\sqrt{2/3} & 1/\sqrt{3} & 0 \cr -1/\sqrt{6} & 1/\sqrt{3} 
& -1/\sqrt{2} \cr -1/\sqrt{6} & 1/\sqrt{3} & 1/\sqrt{2}},
\end{equation}
which is very close to what is experimentally observed. 
However, a specific testable prediction of this idea is so far lacking. 
Recently, a model based on $T_7$ and gauged $B-L$ has been shown~\cite{ckmo11} 
to be testable at the Large Hadron Collider (LHC), but it depends on 
observing the $Z'_{B-L}$ gauge boson, which may be too heavy to be produced.
In this paper, we show that it is not necessary to extend the gauge symmetry 
of the standard model (SM).  All one needs is to find a standard-model-like 
Higgs boson $H^0$ whose decay may reveal the residual $Z_3$ symmetry, i.e. 
lepton flavor triality~\cite{m10}, coming from $A_4$, $T_7$, $\Delta(27)$, and 
possibly other non-Abelian discrete symmetries~\cite{ikoost10}.  If 
kinematically allowed, 
this unusual and easily detectable decay is predicted to be observable at 
the LHC and perhaps at the Tevatron as well. Specifically, $H^0 \to \psi_2^0 
\bar{\psi}_2^0$ should be searched for, where $\psi_2^0$ is a light scalar 
with dominant decays to $\tau^+ \mu^-$ and $\tau^- e^+$, resulting thus for 
example in the easily detectable configuration $H^0 \to 
(\tau^- e^+)(\tau^- \mu^+)$.  (The idea of using the decay of $H^0$ to 
two exotic scalars to discover the underlying flavor symmetry has been 
explored recently~\cite{blp11}, using a previously proposed $S_3$ 
model~\cite{cfm04}.)

In Sec.~II we reiterate how the notion of lepton triality is realized in the 
lepton Higgs Yukawa interactions.  In Sec.~III we analyze the general scalar 
potential of four Higgs electroweak doublets transforming as an irreducible 
triplet plus a singlet of $A_4$, $T_7$, and $\Delta(27)$.  We show that they 
share a common solution which is useful for proving lepton triality 
experimentally.  In Sec.~IV we obtain all the Higgs boson masses in a 
specific scenario which is also consistent with present phenomenological 
bounds.  In Sec.~V we show that the decay $H^0 \to \psi_2 \bar{\psi}_2$ 
has a significant branching fraction for a wide range of $m_H$ values. 
In Sec.~VI we discuss how $H^0$ itself may be observed through lepton 
triality at the Large Hadron Collider and its discovery reach. In Sec.VII  
we have some concluding remarks. 

\section{Charged-lepton Higgs interactions}
 
The first thing to notice is that if $L_i = (\nu,l)_i \sim \underline{3}$, 
$l^c_i \sim \underline{1}_i, i = 1,2,3$, and $\Phi_i = (\phi^+,\phi^0)_i 
\sim \underline{3}$ under $A_4$, $T_7$, or $\Delta(27)$, the Yukawa couplings 
$L_i l^c_j \tilde{\Phi}_k$, where $\tilde{\Phi}_k = (\bar{\phi}^0, -\phi^-)_k$, 
are of the same form, leading to the charged-lepton mass matrix~\cite{mr01}
\begin{equation}
m_l = \pmatrix{y_1 v_1 & y_2 v_1 & y_3 v_1 \cr y_1 v_2 & \omega^2 y_2 v_2 & 
\omega y_3 v_2 \cr y_1 v_3 & \omega y_2 v_3 & \omega^2 y_3 v_3} = 
{1 \over \sqrt{3}} \pmatrix{1 & 1 & 1 \cr 1 & \omega^2 & \omega\cr 1 & \omega 
& \omega^2} \pmatrix{y_1 & 0 & 0 \cr 0 & y_2 & 0 \cr 0 & 0 & y_3} v,
\end{equation}
where $\omega = \exp(2\pi i/3) = -1/2 + i \sqrt{3}/2$, and the condition 
\begin{equation}
v_1 = v_2 = v_3 = v/\sqrt{3}
\end{equation}
has been imposed. Note that this condition is not {\it ad hoc} because it 
corresponds to a residual $Z_3$ symmetry and is thus protected against 
arbitrary corrections.  As first shown~\cite{m04} for $A_4$, then also 
recently~\cite{ckmo11} for $T_7$, this leads naturally to neutrino 
tribimaximal mixing, provided that the neutrino mass matrix has a special 
form, which is realized differently for $A_4$ and $T_7$.  In either case, 
as well as that of $\Delta(27)$, the charged-lepton Higgs interactions are 
completely fixed to be the following:
\begin{eqnarray}
{\cal L}_{int} &=& v^{-1}[m_\tau \bar{L}_\tau \tau_R + m_\mu \bar{L}_\mu 
\mu_R + m_e \bar{L}_e e_R] \phi_0 \nonumber \\ 
&+& v^{-1}[m_\tau \bar{L}_\mu \tau_R + m_\mu \bar{L}_e \mu_R + m_e 
\bar{L}_\tau e_R] \phi_1 \nonumber \\
&+& v^{-1}[m_\tau \bar{L}_e \tau_R + m_\mu \bar{L}_\tau \mu_R + m_e 
\bar{L}_\mu e_R] \phi_2 + H.c.,
\end{eqnarray}
where $v = \langle \phi_0^0 \rangle$ and 
\begin{equation}
\pmatrix{\Phi_1 \cr \Phi_2 \cr \Phi_3} = {1 \over \sqrt{3}} \pmatrix 
{1 & 1 & 1 \cr 1 & \omega^2 & \omega \cr 1 & \omega & \omega^2} 
\pmatrix{\phi_0 \cr \phi_1 \cr \phi_2},
\end{equation}
displaying thus explicitly the important residual $Z_3$ symmetry, i.e. lepton 
triality~\cite{m10}, under which
\begin{equation}
e,~\mu,~\tau \sim 1,~\omega^2,~\omega, ~~~~~\phi_{0,1,2} \sim 1,~\omega,
~\omega^2.
\end{equation}

Whereas $\phi^\pm_{1,2}$ are degenerate in mass, the $\phi^0_{1,2}
(\bar{\phi}^0_{1,2})$ sector is more complicated.  As already shown~\cite{m09}, 
the mass eigenstates here are not $\phi^0_{1,2}$ but rather
\begin{equation}
\psi^0_{1,2} = {1 \over \sqrt{2}} (\phi^0_1 \pm \bar{\phi}^0_2),
\end{equation}
with different masses $m_{1,2}$.  Note that $\phi^0_1 \sim \omega$ and 
$\phi^0_2 \sim \omega^2$, hence $\psi^0_{1,2} \sim \omega$ and 
$\bar{\psi}^0_{1,2} \sim \omega^2$.  As a result of lepton triality, the 
rare decay $l_1^+ \to l_2^+ l_3^+ l_4^-$ allows only two 
possibilities~\cite{m10}
\begin{equation}
\tau^+ \to \mu^+ \mu^+ e^-, ~~~ \tau^+ \to e^+ e^+ \mu^-,
\end{equation}
and the radiative decay $l_1 \to l_2 \gamma$ is not allowed.  The present 
experimental upper limit of the branching fraction of $\tau^+ \to \mu^+ \mu^+ 
e^-$ is $2.3 \times 10^{-8}$, implying thus only the bound
\begin{equation}
{m_1 m_2 \over \sqrt{m_1^2+m_2^2}} > 22~{\rm GeV} \left({174~{\rm GeV} 
\over v} \right).
\end{equation}
On the other hand, the $Z$ gauge boson couples to $\psi_1^0 
\bar{\psi}_2^0 + \psi_2^0 \bar{\psi}_1^0$, i.e. the analog of $AH$ 
in the two-Higgs-doublet model, hence the condition
\begin{equation}
m_1 + m_2 > 209~{\rm GeV}
\end{equation}
also applies.  Otherwise, $e^+ e^- \to Z \to \psi_1^0 \bar{\psi}_2^0 + 
\psi_2^0 \bar{\psi}_1^0$ would have been detected at LEPII which reached 
a peak energy of 209 GeV.  In the following, we will show in detail 
how $m_2$ may be small enough, say 50 GeV, so that $m_H > 2m_2$ and $H^0$ 
will decay into $\psi_2^0 \bar{\psi}_2^0$ and be observed.

\section{Higgs structure in $A_4$, $T_7$, and $\Delta(27)$}
 
For each of the three non-Abelian discrete symmetries $A_4$, $T_7$ and 
$\Delta(27)$, there are four Higgs doublets to be considered: $\eta \sim 
\underline{1}_1$ and $\Phi_{1,2,3} \sim \underline{3}$.  We assume that 
quarks are all singlets, so they couple only to $\eta$, whereas leptons 
transform nontrivially and couple to $\Phi_i$, as already discussed.  The 
quartic scalar potential of $n$ Higgs doublets has in general $n^2(n^2+1)/2$ 
terms. For $n=4$, without any symmetry, there would be 136 terms.  However, 
there are only 10, 7, and 8 terms respectively for $A_4$, $T_7$, and 
$\Delta(27)$.  We will show that a common solution exists for all 3 
cases, involving only 5 quartic couplings, which will provide us with 
the desirable scenario of observable $H^0 \to \psi^0_2 
\bar{\psi}^0_2$ decay.
 
Consider the following quartic Higgs potential:
\begin{eqnarray}
V_4 &=& {1 \over 2} \lambda_0 (\eta^\dagger \eta)^2 + {1 \over 2} \lambda_1 
(\Phi_1^\dagger \Phi_1 + \Phi_2^\dagger \Phi_2 + \Phi_3^\dagger \Phi_3)^2 
\nonumber \\ 
&+&  \lambda_2 |\Phi_1^\dagger \Phi_1 + \omega \Phi_2^\dagger \Phi_2 + 
\omega^2 \Phi_3^\dagger \Phi_3|^2 + \lambda_3 (|\Phi_1^\dagger \Phi_2|^2 + 
|\Phi_2^\dagger \Phi_3|^2 + |\Phi_3^\dagger \Phi_1|^2) \nonumber \\  
&+& {1 \over 2} \lambda_4 [(\Phi_1^\dagger \Phi_2)^2 + (\Phi_2^\dagger \Phi_3)^2 
+ (\Phi_3^\dagger \Phi_1)^2] + H.c. + \lambda_5 |\Phi_1^\dagger \Phi_2 + 
\Phi_2^\dagger \Phi_3 + \Phi_3^\dagger \Phi_1|^2 \nonumber \\ 
&+& \lambda_6 |\Phi_1^\dagger \Phi_2 + \omega \Phi_2^\dagger \Phi_3 + 
\omega^2 \Phi_3^\dagger \Phi_1|^2 + \lambda_7 |\Phi_1^\dagger \Phi_2 + \omega^2 
\Phi_2^\dagger \Phi_3 + \omega \Phi_3^\dagger \Phi_1|^2 \nonumber \\ 
&+& f_1 (\eta^\dagger \eta)(\Phi_1^\dagger \Phi_1 + \Phi_2^\dagger \Phi_2 
+ \Phi_3^\dagger \Phi_3) + f_2 (|\eta^\dagger \Phi_1|^2 + |\eta^\dagger \Phi_2|^2 
+ |\eta^\dagger \Phi_3|^2) \nonumber \\ 
&+& f_3 [(\eta^\dagger \Phi_1)(\Phi^\dagger_2 \Phi_1) + (\eta^\dagger \Phi_2)
(\Phi^\dagger_3 \Phi_2) + (\eta^\dagger \Phi_3)(\Phi^\dagger_1 \Phi_3)] + H.c. 
\nonumber \\ 
&+& {1 \over 2} f_4 [(\eta^\dagger \Phi_1)^2 + (\eta^\dagger \Phi_2)^2 + 
(\eta^\dagger \Phi_3)^2] + H.c. \nonumber \\ 
&+& f_5 [(\eta^\dagger \Phi_1)(\Phi^\dagger_2 \Phi_3) + (\eta^\dagger \Phi_2)
(\Phi^\dagger_3 \Phi_1) + (\eta^\dagger \Phi_3)(\Phi^\dagger_1 \Phi_2)] + H.c. 
\nonumber \\ 
&+& f_6 [(\eta^\dagger \Phi_1)(\Phi^\dagger_3 \Phi_2) + (\eta^\dagger \Phi_2)
(\Phi^\dagger_1 \Phi_3) + (\eta^\dagger \Phi_3)(\Phi^\dagger_2 \Phi_1)] + H.c. 
\end{eqnarray}
For $A_4$, $\lambda_5 = \lambda_6 = \lambda_7 = f_3 = 0$. For $T_7$, 
$\lambda_4 = \lambda_5 = \lambda_6 = \lambda_7 = f_4 = f_5 = f_6 = 0$.
For $\Delta(27)$, $\lambda_3 = \lambda_4 = f_3 = f_4 = f_5 = f_6 = 0$. 
The common terms are then $\lambda_{0,1,2}$ and $f_{1,2}$.  However, 
there is an identity, i.e. $\lambda_5 = \lambda_6 = \lambda_7$ for 
$\Delta(27)$ is equivalent to having the $\lambda_3$ term in $A_4$ and 
$T_7$.  Hence we can look for a desirable solution applicable to all three 
with nonzero values of $\lambda_{0,1,2,3}$ and $f_{1,2}$.  It turns out 
that $f_2=0$ may also be assumed for simplicity, so our following 
analysis involves only five quartic couplings.  Of course, this may not be 
the true structure of the correct (and presumably much more complicated) 
model of lepton flavor symmetry, but it is a starting point to demonstrate 
phenomenologically that this idea can be tested experimentally.

We now rotate to the $\phi_{0,1,2}$ basis using Eq.~(5), anticipating the 
breaking of $A_4$, $T_7$ or $\Delta(27)$ into $Z_3$ with $\langle \phi_0^0 
\rangle \neq 0$, but $\langle \phi_1^0 \rangle = \langle \phi_2^0 \rangle 
= 0$.
\begin{eqnarray}
V_4 &=& {1 \over 2} \lambda_0 (\eta^\dagger \eta)^2 + {1 \over 2} \lambda_1 
(\phi_0^\dagger \phi_0 + \phi_1^\dagger \phi_1 + \phi_2^\dagger \phi_2)^2 
\nonumber \\ 
&+& \lambda_2 |\phi_0^\dagger \phi_1 + \phi_1^\dagger \phi_2 + \phi_2^\dagger 
\phi_0|^2 + {1 \over 3} \lambda_3 (|\phi_0^\dagger \phi_0 + \omega 
\phi_1^\dagger \phi_1 + \omega^2\phi_2^\dagger \phi_2|^2 \nonumber \\ 
&+& |\phi_0^\dagger \phi_1 + \omega \phi_1^\dagger \phi_2 + \omega^2 
\phi_2^\dagger \phi_0|^2 + |\phi_0^\dagger \phi_1 + \omega^2 \phi_1^\dagger 
\phi_2 + \omega \phi_2^\dagger \phi_0|^2) \nonumber \\ 
&+& f_1 (\eta^\dagger \eta)(\phi_0^\dagger \phi_0 + \phi_1^\dagger \phi_1 
+ \phi_2^\dagger \phi_2).
\end{eqnarray}
To this we add the bilinear terms which break $A_4$, $T_7$, or $\Delta(27)$, 
but preserve $Z_3$.
\begin{eqnarray}
V_2 &=& m_0^2(\eta^\dagger \eta) + \mu_0^2 (\phi_0^\dagger \phi_0) 
+ \mu_1^2 (\phi_1^\dagger \phi_1) + \mu_2^2 (\phi_2^\dagger \phi_2) 
+ m_{12}^2 (\eta^\dagger \phi_0) + H.c.
\end{eqnarray}
We note that the non-Abelian discrete symmetry is assumed to be broken both 
spontaneously and explicitly by soft terms.  Without the latter, unwanted 
massless Goldstone bosons may appear and severe constraints on the 
physical masses of the Higgs bosons may result, as discussed in two 
recent studies~\cite{mmp11,abmp11}, where $A_4$ only is considered. 
We now extract the masses of all the physical scalar particles and show 
that our desired scenario is indeed possible for a wide range of 
parameters.  

\section{Higgs boson masses}
 
Let $\langle \eta^0 \rangle = v \cos \beta$ and $\langle \phi^0_0 \rangle 
= v \sin \beta$, where $v = (2 \sqrt{2} G_F)^{-1/2} = 174$ GeV, then the 
two stability conditions for the minimum of $V_2 + V_4$ are given by
\begin{eqnarray}
0 &=& m_0^2 + m_{12}^2 \tan \beta + \lambda_0 v^2 \cos^2 \beta + f_1 v^2 
\sin^2 \beta, \\ 
0 &=& \mu_0^2 + m_{12}^2 \cot \beta + [\lambda_1 + (2/3) \lambda_3] v^2 
\sin^2 \beta + f_1 v^2 \cos^2 \beta,
\end{eqnarray}
The masses of the five physical Higgs bosons in this sector are given by
\begin{eqnarray}
m^2(H^\pm) = m^2(A) = {-m_{12}^2 \over \sin \beta \cos \beta},
\end{eqnarray}
\begin{eqnarray}
m^2(H^0,h^0) = \pmatrix{-m_{12}^2 \tan \beta + 2 \lambda_0 v^2 \cos^2 \beta & 
m_{12}^2 + 2 f_1 v^2 \sin \beta \cos \beta \cr m_{12}^2 + 2 f_1 v^2 \sin \beta 
\cos \beta & -m_{12}^2 \cot \beta + 2 [\lambda_1 +(2/3) \lambda_3] v^2 
\sin^2 \beta}.
\end{eqnarray}
For simplicity, we will assume
\begin{equation}
2 f_1 v^2 = {-m_{12}^2 \over \sin \beta \cos \beta},
\end{equation}
so that $H^0$ does not mix with $h^0$.  We will also assume 
$\sin \beta = \cos \beta = {1/\sqrt{2}}$, then the masses become
\begin{equation}
m^2(H^\pm) = m^2(A) = 2 f_1 v^2, ~~~ m^2(H^0) = (\lambda_0 + f_1) v^2, ~~~ 
m^2(h^0) = \left( \lambda_1 + {2 \over 3}\lambda_3 + f_1 \right) v^2.
\end{equation}
Since $H^0 = \sqrt{2} Re(\eta)$, 
it will couple to quarks as in the standard model, except for the enhanced 
Yukawa coupling by the factor $1/\cos \beta = \sqrt{2}$.  This allows it 
to be produced by the usual one-loop gluon-gluon process at the LHC~\cite{bd11}.

In the $\phi_{1,2}$ sector, we will make another simplifying assumption, 
i.e. $\mu_1^2 = \mu_2^2 = \mu_{12}^2$.  Then their masses are given by
\begin{eqnarray}
m^2(\phi^\pm_{1,2}) &=& \mu_{12}^2 + \left( {1 \over 2}\lambda_1 - {1 \over 6}
\lambda_3 + {1 \over 2}f_1 \right) v^2, \\ 
m^2(\psi^0_1) &=& \mu_{12}^2 + \left( {1 \over 2}\lambda_1 + \lambda_2 + 
{1 \over 2} f_1 \right) v^2, \\ 
m^2(\psi^0_2) &=& \mu_{12}^2 + \left( {1 \over 2}\lambda_1 + {1 \over 3} 
\lambda_3 + {1 \over 2} f_1 \right) v^2. 
\end{eqnarray}
Since $H^+ H^-$, $AH^0$, $Ah^0$, $\phi^+_{1,2} \phi^-_{1,2}$, and 
$\psi^0_{1,2} \bar{\psi}^0_{2,1}$ all couple to the $Z$, their nonobservation 
at LEPII implies
\begin{eqnarray}
&& \sqrt{2f_1} v > 104.5~{\rm GeV}, \\ 
&& (\sqrt{2f_1} + \sqrt{\lambda_0 + f_1}) v > 209~{\rm GeV}, \\ 
&& (\sqrt{2f_1} + \sqrt{\lambda_1 + (2/3) \lambda_3 + f_1}) v > 209~{\rm GeV}, 
\\ 
&& -{1 \over 2} \lambda_3 v^2 = m^2(\phi^\pm_{1,2}) - m^2(\psi^0_2) > 
(104.5~{\rm GeV})^2 - m^2(\psi^0_2), \\ 
&& \left( \lambda_2 - {1 \over 3} \lambda_3 \right) v^2 = m^2(\psi^0_1) - 
m^2(\psi^0_2) > (209~{\rm GeV})(209~{\rm GeV} - 2m(\psi^0_2)).
\end{eqnarray}

\begin{figure}
\includegraphics[scale=0.6]{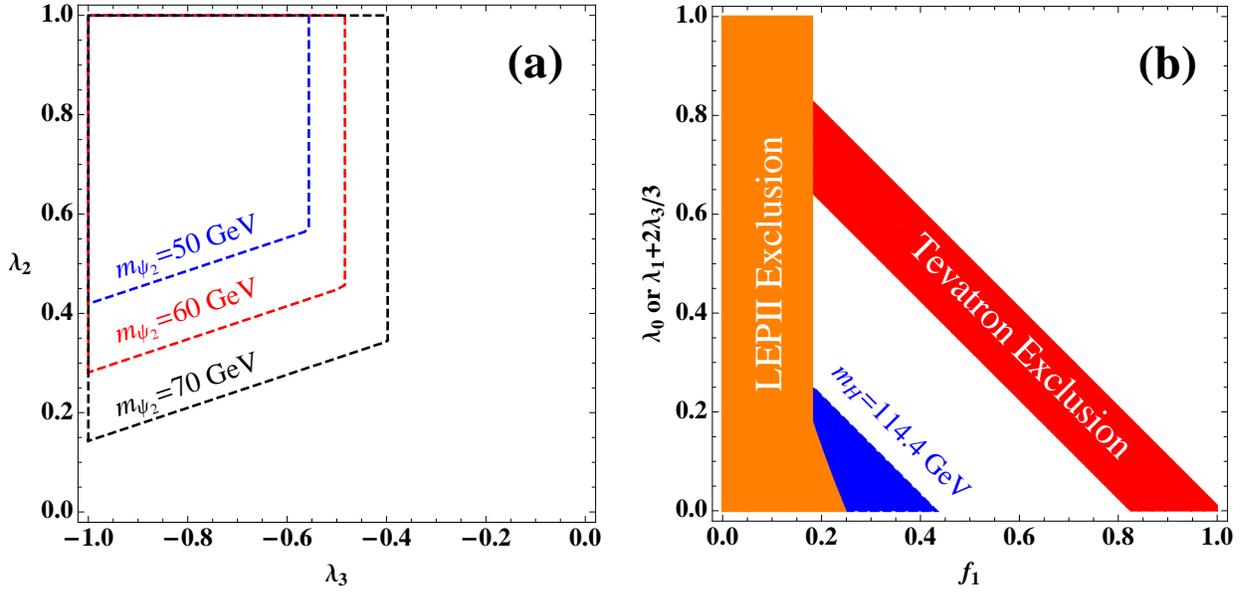}
\caption{(a) Allowed region in the plane of  $\lambda_2$ and $\lambda_3$ for 
$m_{\psi_2} =$ (50, 60, 70) GeV where the region inside each dashed box
is allowed. (b) Allowed region in the plane of $\lambda_0$ (or $\lambda_1 + 
(2/3) \lambda_3$) and $f_1$ where the shaded regions are ruled out by several 
experiments as explained in the text.\label{fig:para}} 
\end{figure}

In Fig.~\ref{fig:para}, we show the allowed region of values for $\lambda_2$ 
and 
$-\lambda_3$, for $m_{\psi_2} = 50$, 60, 70 GeV.  We show also the allowed 
region of values for either $\lambda_0$ or $\lambda_1 + (2/3)\lambda_3$ 
and $f_1$.  The constraints coming from the nonobservation of 
the standard-model Higgs boson at LEPII, i.e.
\begin{equation}
m_{H,h} > 114.4~{\rm GeV},
\end{equation}
as well as the Tevatron exclusion, i.e.~\cite{teva10}
\begin{equation}
158~{\rm GeV} < m_{H,h} < 175~{\rm GeV},
\end{equation}
are also shown.  It is clear that there is a wide range of parameter space 
for our desired scenario. It should of course be added that the analysis 
which obtained these bounds are based on the SM.  Here, $H^0$ has other 
decay modes, so these bounds are not necessarily obeyed.  Thus Fig.~1 is 
merely an illustration that the allowed parameter space for this model is 
not closed.

\section{Higgs boson $H^0$ decay branching fractions}

\begin{figure}
\begin{center}
\includegraphics[scale=0.45]{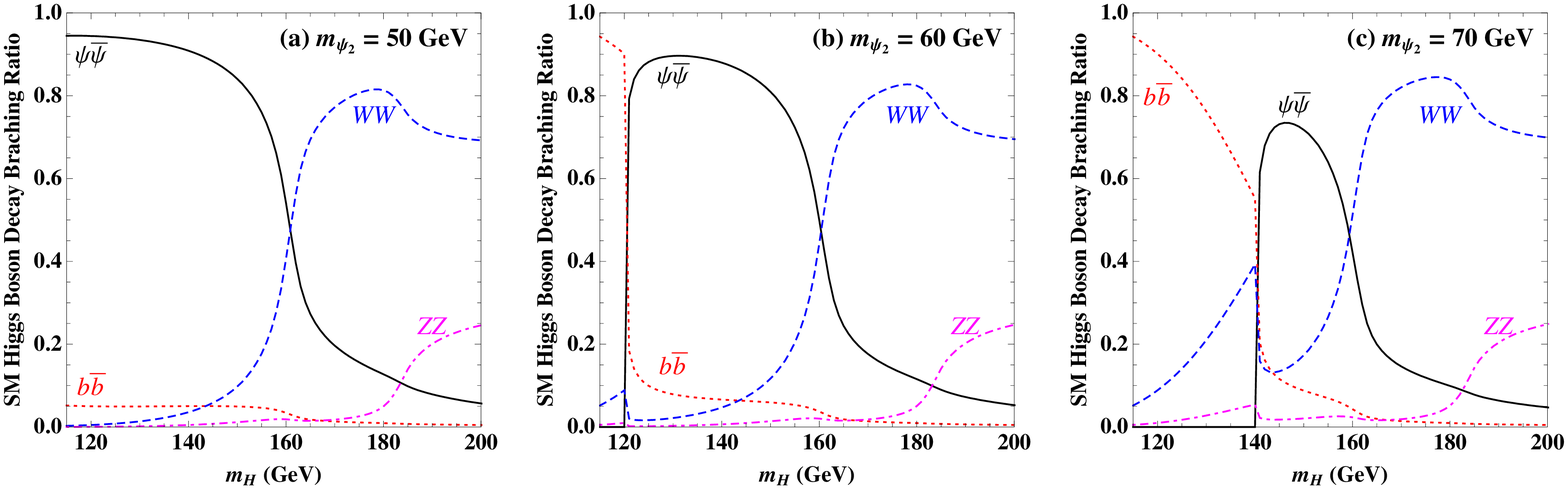}
\includegraphics[scale=0.45]{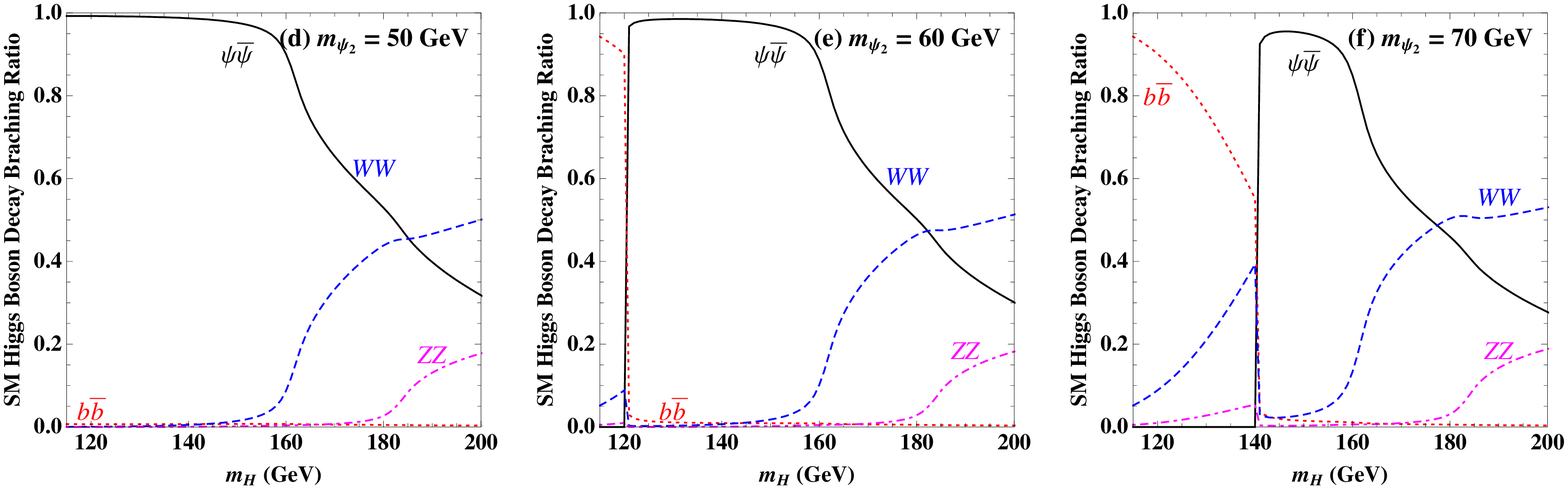}
\caption{Decay branching fractions of the Higgs boson $H^0$ as a function of 
$m_H$ for $m_\psi =$ 50, 60, 70 GeV with $f_1=0.18$ (a,b,c) and 
$f_1=0.5$ (d,e,f).
\label{fig:br}} 
\end{center}
\end{figure}

The production of $H^0$ is similar to that of the standard-model Higgs 
boson.  Since it couples to quarks (in particular the $t$ quark) with 
$\sqrt{2}$ times the standard-model coupling, the gluon-gluon production 
of $H^0$ has 2 times the expected cross section.  Once produced, it will 
decay into the usual channels, such as $b \bar{b}$, $W^-W^+$, $ZZ$, etc. 
However, the $H^0 \to \psi^0_2 \bar{\psi}^0_2$ decay rate is substantial 
if kinematically allowed.  Its coupling is $f_1 v$, hence
\begin{equation}
\Gamma_\psi = {f_1^2 v^2 \over 16 \pi m_H} \sqrt{ 1 - {4m_{\psi_2}^2 
\over m_H^2}},
\label{h2psi}
\end{equation}
whereas below (and above) threshold, there is also a contribution from the 
virtual decay of $\psi_2$ or $\bar{\psi}_2$ to leptons, with a three-body 
decay rate given by
\begin{equation}
\Gamma_{\psi*} = {f_1^2 m_\tau^2 \over 64 \pi^3 m_H} \int^{r^2}_{2r-1} 
{dy (r^2-y) \sqrt{(1+y)^2 - 4r^2} \over y^2 + r^4 \Gamma_2^2/m_{\psi_2}^2},
\end{equation}
where $r = m_{\psi_2}/m_H$ and 
\begin{equation}
\Gamma_2 = {m_\tau^2 m_{\psi_2} \over 8 \pi v^2}
\end{equation}
is the decay width of $\psi_2$.  However, this 3-body contribution is very 
small and can be safely neglected.  The other non-negligible decay modes are 
as in the SM for $H^0 \to WW, ZZ$ and 2 times as large for 
$H^0 \to b \bar{b}$.  We plot in Fig.~\ref{fig:br}
the branching fraction of $H^0 \to \psi_2 \bar{\psi}_2$ as a function 
of $m_H$ from 115 to 200 GeV for $m_2 = 50$, 60, 70 GeV, using  
the \underline{minimum} value of $f_1=0.18$ from Eq.~(23), and also $f_1=0.5$. 
In Table~\ref{hwidth} we list the total widths ($\Gamma$) of $H^0$ 
as well as its branching ratios (Br) to $\psi_2\bar\psi_2$ for five values 
of $m_H$ using $f_1=0.18$ and 0.5. We see that $H^0 \to \psi_2 \bar{\psi}_2$ is 
easily observable in a wide range of  $m_H$ values for $f_1=0.18$ and much 
better for $f_1=0.5$.  
Once $\psi^0_2$ and $\bar{\psi}^0_2$ are produced, 
$\psi^0_2$ will decay equally into $\tau^+ \mu^-$ and $\tau^- e^+$ according 
to Eq.~(4), and $\bar{\psi}^0_2$ into $\tau^- \mu^+$ and $\tau^+ e^-$.  
Thus 25\% of the events will be $(\tau^- \mu^+)(\tau^- e^+)$, an unmistakable 
signature at the LHC.  It also has much less background 
than $b \bar{b}$, which is a serious obstacle to the detection of the 
standard-model Higgs boson at a hadron collider, but not in our scenario.

\begin{table}[t]
\caption{Total width of $H^0$ and its decay branching ratio 
to $\psi^0_2\bar{\psi}^0_2$ for $f_1=0.18$ and 0.5. \label{hwidth}}

\begin{tabular}{|c|cc|cc|cc|cc|cc|cc|}
\hline 
$m_{H}$ & \multicolumn{6}{c|}{$f_1=0.18$} & \multicolumn{6}{c|}{$f_1=0.5$} 
\tabularnewline
\cline{2-13}
(GeV)   & \multicolumn{2}{c|}{$m_{\psi_2}=50~{\rm GeV}$}& \multicolumn{2}{c|}
{$m_{\psi_2}=60~{\rm GeV}$}
        & \multicolumn{2}{c|}{$m_{\psi_2}=70~{\rm GeV}$}& \multicolumn{2}{c|}
{$m_{\psi_2}=50~{\rm GeV}$}
        & \multicolumn{2}{c|}{$m_{\psi_2}=60~{\rm GeV}$}& \multicolumn{2}{c|}
{$m_{\psi_2}=70~{\rm GeV}$}  \tabularnewline
\cline{2-13} 
        & $\Gamma$ & Br & $\Gamma$ & Br& $\Gamma$ & Br& $\Gamma$ & Br& 
$\Gamma$ & Br& $\Gamma$ & Br \tabularnewline
\hline
 110 & 0.080 & 0.942 & 0.005 & 0.000 & 0.005 & 0.000 & 0.588 & 0.996 & 0.005 
& 0.000 & 0.005 & 0.000 \tabularnewline  
 150 & 0.118 & 0.841 & 0.099 & 0.810 & 0.067 & 0.718 & 0.784 & 0.988 & 0.635 
& 0.985 & 0.388 & 0.975 \tabularnewline
 200 & 1.516 & 0.057 & 1.509 & 0.053 & 1.500 & 0.048 & 2.096 & 0.478 & 2.045 
& 0.462 & 1.979 & 0.431 \tabularnewline
 250 & 4.123 & 0.018 & 4.120 & 0.017 & 4.116 & 0.016 & 4.614 & 0.219 & 4.590 
& 0.210 & 4.560 & 0.200 \tabularnewline
 300 & 8.571 & 0.007 & 8.569 & 0.007 & 8.567 & 0.007 & 8.993 & 0.102 & 8.979 
& 0.099 & 8.963 & 0.096 \tabularnewline
\hline
\end{tabular}
\end{table}

\section{Collider phenomenology at 7~TeV}

\subsection{Discovery potential}

We now study in detail the process $gg\to H^0 \to \psi_2 \bar{\psi}_2$ with 
the subsequent decay $\psi_2 \to \tau^- e^+$ and $\bar{\psi}_2 \to \tau^- 
\mu^+$ at the LHC with $E_{\rm cm}=7~{\rm TeV}$. The collider signature of 
interest is 
\be
e^+\mu^+\ell^-\ell^- + \met,
\ee
where $\ell=e,\mu$ and the missing transverse energy ($\met$) originates from 
the unobserved neutrinos from the two $\tau$ decays.  The dominant backgrounds 
yielding the same signature are the processes (generated
by MadEvent/MadGraph~\cite{Alwall:2007st}):
\bea
ZZ		 &:& pp \to ZZ, Z\to \ell^+\ell^-, Z\to \tau^+\tau^-, \tau^\pm 
\to \ell^\pm \nu \bar{\nu} \nonumber \\
WWZ		 &:& pp \to W^+ W^- Z, W^\pm \to \ell^\pm \nu, Z\to 
\ell^+\ell^-, \nonumber \\
t\bar{t} &:& pp \to t\bar{t} \to b(\to \ell^-) \bar{b}(\to \ell^+) W^+ W^-, 
W^\pm \to \ell^\pm \nu, \nonumber \\
Zb\bar{b}&:& pp \to Zb(\to \ell^-) \bar{b}(\to \ell^+), Z\to \ell^+\ell^-. 
\nonumber
\eea
We require no jet tagging and focus on only events with both $e^+$ and 
$\mu^+$ in the final state. The first two processes are the irreducible 
background, while the last two are reducible as they only contribute when 
some observable particles escape detection, carrying away small transverse 
momentum ($p_T$) or falling out of the detector rapidity coverage. 

In our analysis, all events are required to pass the following basic 
acceptance cuts:
\bea
&& n_\ell=4, \qquad n_{e^+}=1,\qquad n_{\mu^+}=1,  \qquad n_{\ell^-}=2\nonumber \\
&& p_T(e^+,\mu^+) > 15~{\rm GeV}, \quad p_T(\ell^-) > 10~{\rm GeV}, \quad
 \met > 15~{\rm GeV}, \nonumber \\
&& \left|\eta_{\ell}\right| < 2.5, \quad \Delta R_{\ell\ell^\prime}\geq 0.4,
\eea
where $\Delta R_{ij}$ is the separation in the plane spanned by the azimuthal 
angle ($\phi$) and the pseudorapidity ($\eta$) between $i$ and $j$, defined as 
\be
\Delta R_{ij}\equiv \sqrt{(\eta_i-\eta_j)^2+(\phi_i-\phi_j)^2}.
\ee
We also model detector resolution effects by smearing the final-state lepton 
energies using
\be
\frac{\delta E}{E} =\frac{10\%}{\sqrt{E/{\rm GeV}}}\oplus 0.7\%. 
\ee
Note that a soft cut is imposed on the negatively charged leptons because 
they originate from $\tau$ decay.  Since the Higgs boson decays predominantly 
into the $\psi_2\bar{\psi}_2$ pair only just above the threshold region, 
these scalars are not boosted.  The leptons from their decays would then  
exhibit only small $p_T$.  The charged lepton from the subsequent $\tau$ 
decay becomes even softer. 

\begin{figure}
\begin{center}
\includegraphics[scale=0.6]{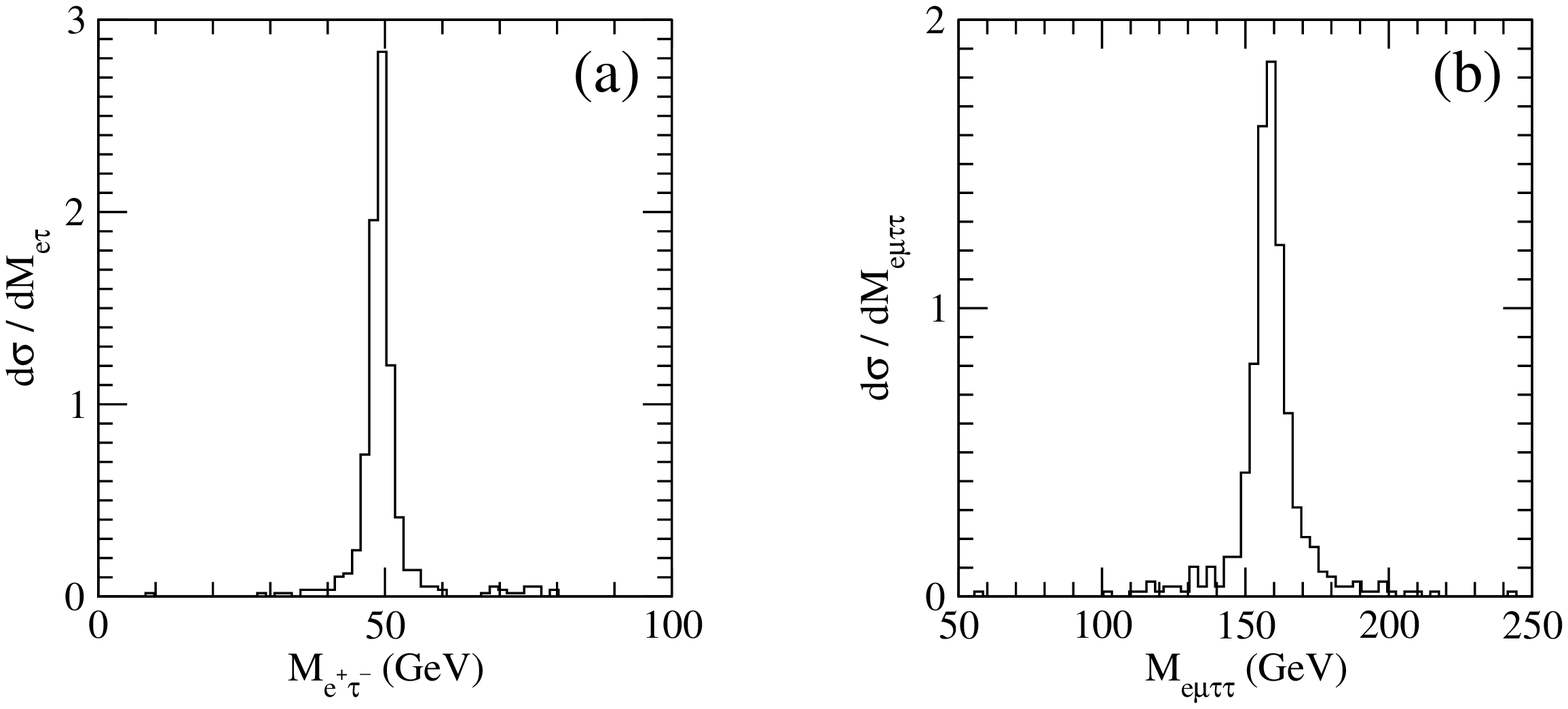}
\caption{(a) Reconstructed $m_{\psi_2}$ from $e^+\tau^-$ and (b) $m_H$ from 
$e^+\mu^+\tau^-\tau^-$ for $m_{\psi_2} = 50$ GeV and $m_H = 160$ GeV.
\label{fig:dist}} 
\end{center}
\end{figure}

To reconstruct the scalar $\psi$, we adopt the collinear approximation 
that the charged lepton and neutrinos from $\tau$ decays are parallel 
due to the large boost of the $\tau$. Such a condition is satisfied to an 
excellent degree because the $\tau$ leptons originate from a heavy scalar 
decay in the signal event.  Denoting by $x_{\tau_i}$ the fraction of the parent 
$\tau$ energy which each observable decay particle carries, the transverse 
momentum vectors are related by~\cite{Rainwater:1998kj}
\be
\vec{\not{\!E}}_T = \left(\frac{1}{x_{\tau_1}} - 1 \right)\vec{p}_1
+\left(\frac{1}{x_{\tau_2}}-1\right)\vec{p}_2. \label{eq:taurec}
\ee
When the decay products are not back-to-back, Eq.~(\ref{eq:taurec}) gives two 
conditions for $x_{\tau_i}$ with the $\tau$ momenta as $\vec{p}_1/x_{\tau_1}$ 
and $\vec{p}_{2}/x_{\tau_2}$, respectively. We further require the calculated 
$x_{\tau_i}$ to be positive to remove the unphysical solutions.
There are two possible combinations of $e^+\ell^-$ clusters 
for reconstructing the scalar $\psi$ and Higgs boson. To 
choose the correct combination, we require the $e^+\ell^-$ pairing to be 
such that $\Delta R_{e^+\ell^-}$ is minimized.
The mass spectra of the reconstructed $\psi$ 
and Higgs boson are plotted in 
Fig.~\ref{fig:dist}(a) and (b), respectively, which clearly 
display sharp peaks around $m_\psi$ and $m_H$.  

\begin{table}
\caption{Cross sections (fb) of signal and SM backgrounds before and after 
cuts, using $f_{1}=0.5$, for five values of $m_{H}$ (GeV) and three values 
of $\psi_2$ mass ($m_{\psi_2}$) after the restriction to 
$e^{+}\mu^{+}\ell^{-}\ell^{-}$
and with tagging efficiencies included. \label{lhc_signal}}
\begin{tabular}{|c|ccc|ccc|ccc|c|ccc|}
\hline 
$m_{H}$ & \multicolumn{3}{c|}{$m_{\psi_2}=50\,{\rm GeV}$} & \multicolumn{3}{c|}
{$m_{\psi_2}=60\,{\rm GeV}$} & \multicolumn{3}{c|}{$m_{\psi_2}=70\,{\rm GeV}$} &
 \multicolumn{4}{c|}{SM backgrounds}
\tabularnewline
\cline{2-14} 
(GeV) & no cut & basic & $x_{i}>0$ & no cut & basic & $x_{i}>0$ &no cut & 
basic & $x_{i}>0$ & & no cut & basic & ${x_{i}}>0$\tabularnewline
\hline
110 &  428.3 & 11.99 & 11.91 & 3.71  & 0.26 & 0.25   & 0.80  & 0.09  & 0.09  
&$t\bar{t}$  & 0.21  & 0.14 & 0.04  \tabularnewline
150 &  216.4 & 8.22  & 8.18  &216.47 & 13.51 & 13.42 & 214.1 & 21.46 & 21.33 
&$ZZ$        & 10.14 & 0.12 & 0.09  \tabularnewline
200 &  54.09 & 4.65  & 4.60  &52.23  & 4.94  & 4.87  & 47.98 & 5.99  & 5.94  
&$Zb\bar{b}$ & 0.83  & 0.13 & 0.06  \tabularnewline
250 &  14.82 & 2.17  & 2.13  &14.37  & 2.17  & 2.14  & 13.81 & 2.38  & 2.35  
&$WWZ$       & 0.06  & 0.03 & 0.01  \tabularnewline
300 &  4.90  & 0.92  & 0.90  &4.70   & 0.92  & 0.90  & 4.46  & 0.93  & 0.92  
&            &       &      &       \tabularnewline
\hline
\end{tabular}
\end{table}%

In Table~\ref{lhc_signal} we show the signal and background cross 
sections (in fb units) before and after our cuts, with $f_1=0.5$, for ten 
values of $m_H$ and three values of $m_\psi$.  Due to the narrow-width 
approximation, the signal process can be factorized, i.e.  
as the simple product of the production of $H$ and its decay as follows:
\be
\sigma(gg\to H \to \psi_2\bar{\psi}_2) = \sigma(gg\to H)\times {\rm Br}
(H\to\psi_2\bar{\psi}_2),
\ee 
where 
\be
{\rm Br}(H\to \psi_2\bar{\psi}_2) \approx \frac{\Gamma_{\psi_2}}{\Gamma_{\psi_2} 
+ \Gamma_{\rm SM}}.
\ee
Since $\Gamma_{\psi_2} \propto f_1^2$, one can extract the corresponding 
signal cross section for values of $f_1$ other than 0.5 (those displayed 
in Teble I) easily from Tables~\ref{hwidth} and \ref{lhc_signal} and 
Eq.~\ref{h2psi}.

\begin{figure}
\begin{center}
\includegraphics[scale=0.5]{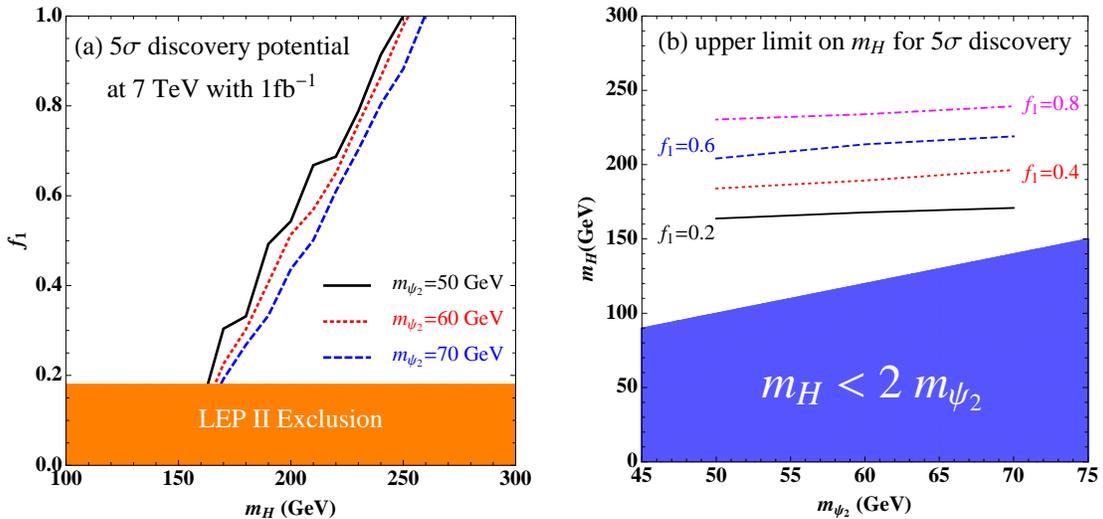}
\caption{Discovery potential of signal (a) in the plane of $f_1$ and $m_H$ 
where the region above each curve is good for a $5\sigma$ discovery, and (b) 
in the plane of $m_H$ and $m_{\psi_2}$.}
\label{lhcfig} 
\end{center}
\end{figure}

In Fig.~\ref{lhcfig} we display the discovery potential of the signal process 
in the plane of $f_1$ and $m_H$ as well as $m_H$ and $m_{\psi_2}$ at the LHC for 
$E_{\rm cm} = 7$ TeV with an integrated luminosity of $1~{\rm fb}^{-1}$. 
Since there is no background after all cuts, one can claim a $5\sigma$ 
discovery once 5 signal events are observed. 

%
%  Higgs search in the WW mode
%
\begin{figure}
\begin{center}
\includegraphics[scale=0.6]{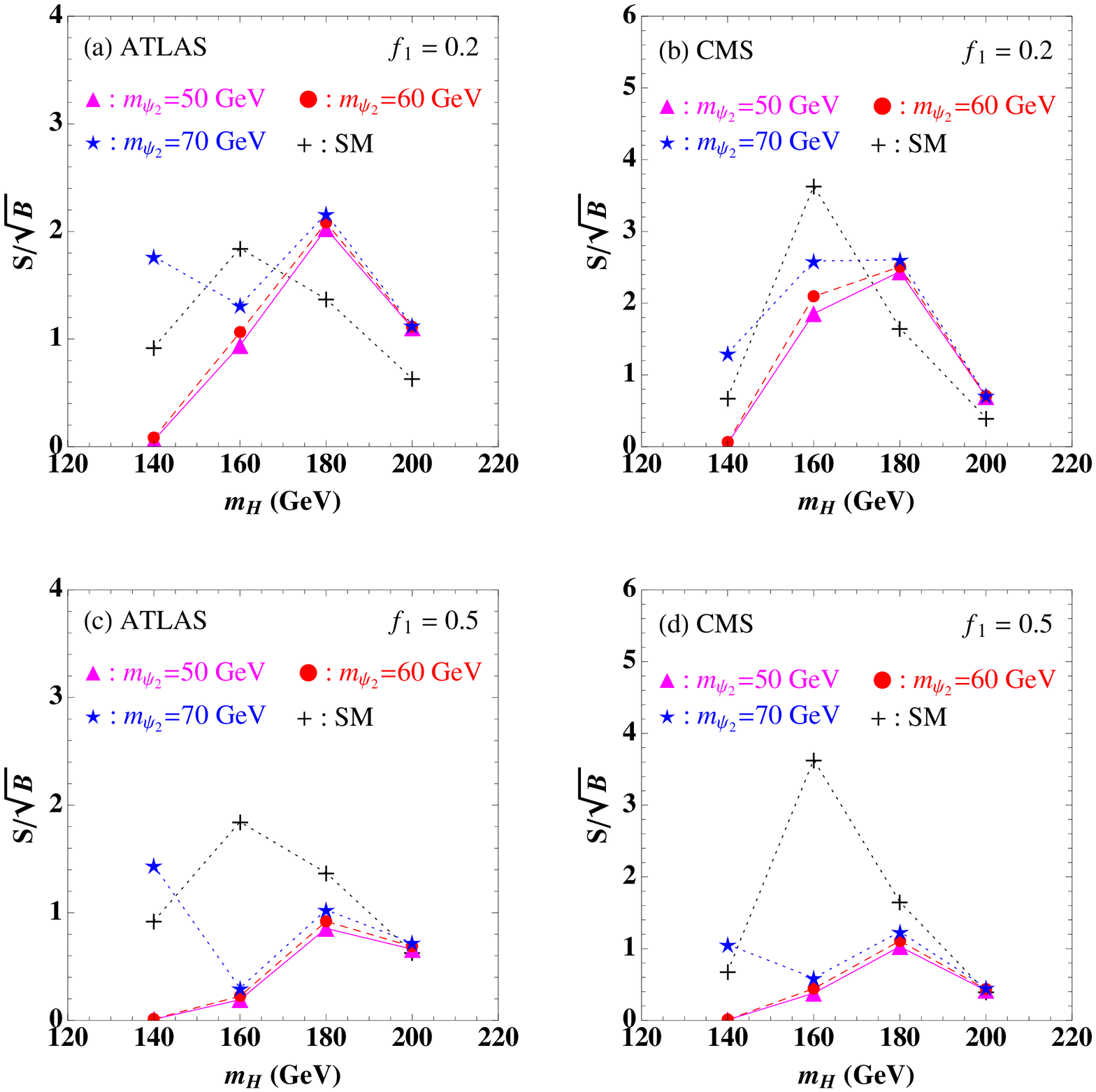}
\caption{Discovery potential of $H$ compared to the SM Higgs boson in the 
$WW$ mode at 7~TeV with an integrated luminosity of $1~{\rm fb}^{-1}$: 
(a) ATLAS and (b) CMS for $f_1=02$; (c) ATLAS and (d) CMS for $f_1=0.5$.}
\label{smhww} 
\end{center}
\end{figure}

\subsection{Impact on the SM Higgs search in $WW$ mode}

The cross section of $H$ production via gluon-gluon fusion is doubled 
because the Yukawa coupling of $H$ to the top quark is enhanced by a factor 
of $\sqrt{2}$. Below we explore the impact of the new decay channel 
$H\to \psi_2\bar{\psi}_2$ on the SM Higgs boson search.
In Ref.~\cite{Berger:2010nc} the SM Higgs discovery potential at 7~TeV 
in the $WW$ mode was studied in detail.  Compared to that, the discovery 
potential of $H$ in our model can be extracted easily via the following 
relation:
\be
\frac{\mathcal{S}}{\mathcal{S}_{SM}}=
\frac{\sigma(gg\to H) \times {\rm Br}(H\to WW)}{\sigma(gg\to H)_{SM}\times 
{\rm Br}(H\to WW)_{SM}}
=2\times \frac{\Gamma_{\rm SM}}{\Gamma_{\rm SM}+\Gamma(H\to 
\psi_2\bar{\psi}_2)}.
\ee
In Fig.~\ref{smhww} we display the discovery significance $S/\sqrt{B}$
at 7~TeV with an integrated luminosity of $1~{\rm fb}^{-1}$ for the ATLAS (a) 
and CMS (b) detectors with $f_1=0.2$ and also with $f_1=0.5$ for ATLAS (c) 
and CMS (d). 
If the $\psi_2\bar{\psi}_2$ mode is forbidden by kinematics, i.e. 
$m_H < 2 m_{\psi_2}$, the SM Higgs search in the $WW$ mode is unaffected. 
However, once the $\psi_2\bar{\psi}_2$ channel is open, the discovery 
potential of the SM Higgs boson in the $WW$ mode is significantly lowered.
For large values of $f_1$, the $WW$ mode is so much suppressed that it will 
be difficult to discover $H$ in this conventional way.

\section{Conclusion}
We have shown that the routine search of the standard-model 
Higgs boson at the LHC may reveal more than just the standard model.  It may 
show evidence of the underlying $Z_3$ lepton flavor symmetry predicted by 
non-Abelian discrete symmetries, such as $A_4$, $T_7$, and $\Delta(27)$, 
which explain successfully the observed pattern of neutrino tribimaximal 
mixing.  The key is the possible decay $H^0 \to \psi^0_2 \bar{\psi}^0_2$ 
with the unusual and easily detectable $(\tau^- \mu^+)(\tau^- e^+)$ final 
state.  In a specific and much simplified scenario, we show that a 
$5\sigma$ discovery is 
possible at the LHC with 1 fb$^{-1}$ for $E_{\rm cm} = 7$ TeV, up to 
$m_H \sim 200$ GeV.  We show that the conventional $WW$ mode in 
the search for the SM Higgs boson may be impacted significantly as well.

\begin{acknowledgments}
Q.H.C. is supported in part by the U.S. DOE under Grant No.~DE-AC02-06CH11357 
and the Argonne National Laboratory and University of Chicago Joint Theory 
Institute Grant 03921-07-137.
A.D. is supported by the Program 
of Academic Recharging C (PAR-C), Ministry of National Education, Indonesia.
E.M. and D.W. are supported in part by the U.~S.~Department of Energy under
Grant No. DE-FG03-94ER40837.
A.D. thanks the Department of Physics and Astronomy, University of California, 
Riverside for hospitality during his visit.   
\end{acknowledgments}

%\newpage
\baselineskip 16pt
\bibliographystyle{unsrt}

\end{document}